\documentclass[preprint,12pt]{aastex}
\usepackage{epstopdf}
\begin{document}

\def\putplot#1#2#3#4#5#6#7{\begin{centering} \leavevmode
\vbox to#2{\rule{0pt}{#2}}
\includegraphics{#1}

\end{centering}}

\def\Msun{M_\odot}
\def\Lsun{L_\odot}
\def\Rsun{R_\odot}

\slugcomment{Submitted to AJ}

\shorttitle{AT Cnc Shell}
\shortauthors{Shara et al}

\title{AT Cnc: A Second Dwarf Nova with a Classical Nova Shell}

\author{Michael~M.~Shara\altaffilmark{1,2}, Trisha~Mizusawa\altaffilmark{1}, Peter Wehinger\altaffilmark{3}, David~Zurek\altaffilmark{1,2,4}, Christopher D. Martin\altaffilmark{5}, James D. Neill\altaffilmark{5}, Karl Forster\altaffilmark{5}  and Mark Seibert\altaffilmark{6}}


\altaffiltext{1}{Department of Astrophysics, American Museum of Natural
History, Central Park West and 79th street, New York, NY 10024-5192}
\altaffiltext{2} {Visiting Astronomer, Kitt Peak National Observatory, National Optical Astronomy Observatory, which is operated by the Association of Universities for Research in Astronomy (AURA) under cooperative agreement with the National Science Foundation. }
\altaffiltext{3} {Steward Observatory, the University of Arizona, 933 North Cherry Avenue, Tucson, AZ 85721}
\altaffiltext{4} {Visiting Astronomer, Steward Observatory, the University of Arizona, 933 North Cherry Avenue, Tucson, AZ 85721}
\altaffiltext{5}{Department of Physics, Math and Astronomy, California Institute of Technology, 1200 East California Boulevard, Mail Code 405-47, Pasadena, California 91125}
\altaffiltext{6}{Observatories of the Carnegie Institution of Washington, 813 Santa Barbara Street, Pasadena, California 91101} 

\begin{abstract}
We are systematically surveying all known and suspected Z Cam-type dwarf novae for classical nova shells. This survey is motivated by the discovery of the largest known classical nova shell, which surrounds the archetypal dwarf nova Z Camelopardalis. The Z Cam shell demonstrates that at least some dwarf novae must have undergone classical nova eruptions in the past, and that at least some classical novae become dwarf novae long after their nova thermonuclear outbursts, in accord with the hibernation scenario of cataclysmic binaries.  Here we report the detection of a fragmented "shell", 3 arcmin in diameter, surrounding the dwarf nova AT Cancri. This second discovery demonstrates that nova shells surrounding Z Cam-type dwarf novae cannot be very rare. The shell geometry is suggestive of bipolar, conical ejection seen nearly pole-on. A spectrum of the brightest AT Cnc shell knot is similar to that of the ejecta of the classical nova GK Per, and of Z Cam, dominated by [NII] emission. Galex FUV imagery reveals a similar-sized, FUV-emitting shell. We determine a distance of 460 pc to AT Cnc, and an upper limit to its ejecta mass of $\sim  5$ $\times 10^{-5} M_\sun$, typical of classical novae. 

\end{abstract}


\section{Introduction and Motivation}

Dwarf and classical novae are all close binary stars, wherein a white dwarf accretes hydrogen-rich matter from its Roche-lobe filling companion, or from the wind of a nearby giant. In dwarf novae, an instability \citep{osa74} episodically dumps much of the accretion disk onto the white dwarf. The liberation of gravitational potential energy then brightens these systems by up to 100-fold every few weeks or months \citep{war95}. This accretion process in dwarf novae must inevitably build an electron degenerate, hydrogen-rich envelope on the white dwarf \citep{sha86}. Theory and detailed simulations predict that once the accreted mass M$_{env}$ reaches of order $10^{-5}M_\odot$, a thermonuclear runaway (TNR) will occur in the degenerate layer of accreted hydrogen. The TNR causes the rapid rise to $\sim 10^{5} L_\sun$ or more, and the high-speed ejection of the accreted envelope (\citet{sha89} and \citet{yar05}) in a classical nova explosion. Theory thus predicts that dwarf novae must inevitably give rise to classical novae.

\citet{col09} have updated the seminal work of \citet{rob75}, finding no evidence for dwarf nova eruptions in the progenitors of classical novae during the decades before the nova explosions. The identified progenitors of almost all classical novae are, instead, novalike variables, in which the mass transfer rate $\dot{M}$ through the accretion disk is too high to permit the disk instability that drives dwarf nova eruptions. This apparent contradiction with theory is explained by the hibernation scenario of cataclysmic variables \citep{sha86} as follows. 

During the millenia before a nova eruption, gravitational radiation drives the white and red dwarfs closer together, enhancing Roche lobe overflow and $ \dot{M}$. The increasingly high mass transfer rate turns a dwarf nova (DN) into a novalike variable centuries before the envelope mass reaches the value needed for a TNR. The higher $\dot{M}$ of a novalike variable chokes off dwarf nova eruptions, hence none are seen as nova progenitors. During the few centuries after a nova eruption the mass transfer rate remains high (due to irradiation of the red dwarf), which again prevents dwarf nova eruptions. A few centuries after a nova eruption, the hibernation scenario predicts that dwarf nova eruptions should begin anew. This is because irradiation of the red dwarf by the cooling white dwarf drops, as does $\dot{M}$. These newly reborn DN  will be the highest mass transfer rate dwarf novae - the Z Camelopardalis stars. Thus within the context of the hibernation scenario one expects old novae to evolve from novalike variables into Z Cam stars in the centuries after nova eruptions. Only these Z Cam stars will be surrounded by old nova shells. As gravitational radiation eventually drives the two stars in a CV together, one expects Z Cam stars to be the most likely progenitors of the novalike variables before they erupt as classical novae. These Z Cam stars will not be surrounded by old nova shells - their shells dispersed many millenia ago. The hibernation scenario thus predicts that some, but not all Z Cam stars should be surrounded by old nova shells.

In 2007 we reported the discovery of a classical nova shell surrounding the prototypical dwarf nova Z Camelopardalis \citep{sha07}. This shell is an order of magnitude more extended than those detected around any other classical nova.The derived shell mass matches that of classical novae, and is inconsistent with the mass expected from a dwarf nova wind or a planetary nebula. The Z Cam shell observationally linked, for the first time, a prototypical dwarf nova with an ancient nova eruption and the classical nova process. This was the first-ever confirmation of a key prediction of cataclysmic binary TNR theory: the accreting white dwarfs in dwarf novae must eventually erupt as classical novae.

Motivated by this discovery, we have been searching for other nova shells surrounding dwarf novae. One of our targets was the Z Cam-like dwarf nova AT Cancri. In a study of AT Cnc, \citet{bon74} found shallow, broad absorption lines, and suggested that the star is an eclipsing binary composed of a DA white dwarf and a faint red dwarf companion. \citet{nog99} found the orbital period to be 0.2011 days,  and detected P Cygni profiles in the asymmetric $H\alpha$ line. A summary of AT Cnc's properties, as well as spectroscopic and photometric observations are given by \citet{nog99}. 

Our optical narrowband imaging of AT Cnc immediately revealed fragmented rings surrounding the star. Followup observations with GALEX confirmed the presence of FUV - emitting material surrounding this dwarf nova.

In Section 2 we describe our observations. We show optical narrowband imagery of the rings of material surrounding AT Cancri in Section 3, and a spectrum of the shell material in section 4. GALEX ultraviolet imagery is presented in section 5. We determine the distance to AT Cnc, and an upper limit to its ejecta mass in section 6. The age of the AT Cnc ejecta is discussed in section 7. The implications of the existence of the ejecta are considered in Section 8, and we briefly summarize our results in Section 9.  
  
\section {Observations and Image Processing}

Narrowband images of AT Cnc in the lines of H$\alpha$ and [NII], and broadband R images  were obtained with the 90Prime camera \citep{wil04} of the 2.3 meter Steward Observatory telescope on 11 November 2007. The camera's focal plane array is populated with a mosaic of four thinned Lockheed 4096 x 4096 pixel CCDs. The camera provides a plate scale of 0.45 arcsec per pixel and a total field-of-view of 1.16 degree x 1.16 degree.The R images' total exposure time was 1800 seconds, while the H$\alpha$ +  [NII] images totaled 5400 seconds. Followup imagery in the same filters was obtained with the Mosaic CCD camera at the prime focus of the Kitt Peak National Observatory Mayall 4 meter telescope. The Mosaic camera on the 4 meter telescope has eight 2048 x 4096 SITe thinned CCDs, and an image scale of 0.26 arcsec/pixel.  Imaging was carried out on the nights of 07 and 09 February 2010, and conditions were generally clear. 18 R band images, each of 180 seconds duration (3240 seconds total exposure) and 18 H$\alpha$ +[NII] images of 1800 seconds each (32400 seconds exposure) were obtained. Images were dithered over both nights during each epoch.

After flatfielding and de-biassing, standalone Daophot \citep{stn87} was used to align the images on each chip; then all of the chips were matched together.  All of the continuum (hereafter "R") and all of the narrowband (hereafter "[NII]") images of each epoch were combined to create the deepest possible image.  The images were stitched together using montage2, a mosaicking program within the standalone Daophot.  After this process was completed individually for both the [NII] and R band images, the narrow and broadband images were matched up with Daophot (which uses triangular stellar patterns for its matching algorithm).

Spectra of the brightest knots in the AT Cnc ejecta were obtained with the R-C spectrograph of the Kitt Peak National Observatory Mayall 4 meter telescope.  A 158 l/mm grating was used because of the faintness of the nebulosity. We combined, using IRAF imcombine, four images (1 x 600 sec plus 3 x 1800 sec, for total of 6000 sec) for the first slit position and two images (1800 sec x 2, for a total of 3600 sec) for the second slit position.  We extracted spectra of each knot using the IRAF task apall.  

Ultraviolet imagery was also obtained with the NASA GALEX satellite. The GALEX image data include 
far-UV (FUV; $\lambda_{eff}$=1516~\AA, $\Delta\lambda$=256~\AA) and near-UV 
(NUV; $\lambda_{eff}$=2267~\AA, $\Delta\lambda$=730~\AA) images in circular fields of diameter
$1\fdg2$. The total exposure in the FUV filter is 1600 sec while that in the NUV filter is 12100 sec.
The spatial resolution is $\sim$5". Details of the GALEX
instrument and data characteristics can be found in \citet{mar05} 
and \citet{mor05}. The imaging data have been
processed under the standard GALEX survey pipeline. 

\section {Imaging of the AT Cnc Shell}

The resulting R and net H$\alpha$+[NII] (narrowband minus R) images from the KPNO 4 meter telescope, taken in 2010 are shown in Figures 1 and 2, respectively. At Cnc is circled in both images. This is one of the deepest narrowband-broadband image pairs ever taken of any nova ejecta. The net narrowband image is dominated by the striking arcs running from the NE through N, W and S of the central star. The two arcs, and their geometry are reminiscent of the hourglass-shaped nebulae of the LMC supernova SN87A \citep{law00} and the planetary nebula  MyCn 18 \citep{sah99}. The near coincidence of the two rings suggests that we are viewing the hourglass almost along its long symmetry axis. Both the rings are extremely fragmented, like the central ring of SN87A seen a decade and more after the supernova eruption. This morphology is suggestive of a fast wind colliding with slow or stationary ejecta from previous outbursts. The shock-dominated spectrum of the brightest knots (next section) supports this interpretation.

\section {Spectrum of the AT Cnc Ejecta}

We placed the 4-meter spectrograph slit at two orientations and positions in order to get spectra of the two brightest knots in AT Cnc's shell.  The first slit, at an angle of 6.6 degrees from North, spans two bright knots very close to each other.  The second slit, at an angle of 119.9 degrees from North, spans a single bright knot.  The slit positions can be seen overlaid on the image of AT Cnc and its shell in Figure 2. Only the first (6000 sec spectrum) yielded sufficient signal to clearly identify the primary emission lines. That spectrum is shown in Figure 3, which is dominated by the emission lines of [NII], [OIII] and [OII]. 

The presence of the [OII] places an upper limit on the density of the emitting gas of about 3000 $cm^{-3}$ \citep{app88}. Unfortunately our spectral resolution is too low to resolve the [OII] doublet and further constrain the density. As we show in section 6, AT Cnc's luminosity (roughly 1 L$_\odot$) and effective temperature (about 10 kKelvin) are too low to photoionize the arcs of ejecta seen in Figure 2. The lack of Balmer lines and the presence of strong [NII] lines (see figure 4) suggests a shock temperature in excess of 20 kKelvins. The emission lines and their ratios are reminiscent of the spectra of the ejecta of the classical nova GK Per (nova 1901)\citep{szd12} and the recurrent nova T Pyx \citep{con97}. Both these are shock ionized due to the collision of rapidly outflowing ejecta with slower moving matter. 

\section{Galex UV Imagery}

The Galaxy Explorer (GALEX) satellite's individual FUV and NUV images of AT Cnc are seen in Figure 5. The small "halo" seen surrounding AT Cnc in the NUV image is instrumental (and seen surrounding other stars of similar brightness), but the faint, extended FUV emission is not. The FUV emission corresponds to the 3 arcmin diameter shells seen in Figure 2. In Figure 6 we compare the optical narrowband [NII] image with the stretched GALEX FUV and NUV images to demonstrate this correspondence. 

\section{The distance to AC Cnc and its Ejected Mass}

Knowing the current size of the AT Cnc ejecta and the range of observed ejection velocities for novae allows us to set upper and lower limits to the time since AT Cnc's last nova eruption. Translating the angular size (3 arcmin in diameter) into a linear size requires knowing the distance to AT Cnc. There is no published parallax or spectroscopic distance for AT Cnc, so we use the infrared Period-Luminosity-Color (PLC) relation derived by \citet{akk07}. This, in turn, demands a knowledge of the reddening to the system. \citet{bru94} note that E(B-V) = 0.0 for  AC Cnc, YZ Cnc and SY Cnc, all of which are located close to Galactic latitude b = + 30 degrees. We adopt the same value for AT Cnc (at b= + 31 degrees).  The distance predicted by the \citet{akk07} PLC method for AT Cnc is then 460 pc. The 1.5 arcmin radius of the AT Cnc ejecta corresponds to 0.2 pc at that distance, and the system luminosity at maximum brightness is thus $\sim  L_\sun$. The spectrum of AT Cnc is dominated by the Balmer lines of the accretion disk; HeI is quite weak, and no trace is seen of HeII. (The modest effective temperature and luminosity of AT Cnc are the basis of our claim in section 4 above that the luminous ejecta of AT Cnc cannot be photo-ionized by the central star).

We can place a very rough upper limit on the ejecta mass as follows. Roughly 100 "blobs" are seen in the emitting rings of Figure 2.
Most of these blobs are unresolved, so we adopt an upper size limit of 1 arcsec for each one.  This corresponds to a physical diameter of 
$\sim 7 \times 10^{13} $ m. Assuming that each blob is spherical with a density less than 3000 $cm^{-3}$ yields an upper limit to the mass in blobs of $\sim  5 \times 10^{-5} M_\sun$, in excellent agreement with theoretical predictions of nova ejecta masses \citep{yar05}.

\section {When did AT Cnc last Erupt as a Nova?} 

Neither the encyclopedic summary of pre-telescopic transient stars of \citet{hop62} nor the carefully culled list of Far Eastern observations of classical novae before 1800 AD of \citet{ste87} contain a classical nova candidate close to the position of AT Cnc. This is in contrast to Z Cam, the oldest classical nova ever recovered. The dynamics of Z Cam's ejecta constrain its last eruption to have occurred more than 1300 years ago \citep{sha12}, and Z Cam's position is consistent with that of the Chinese nova of 77 BCE  \citep{joh07}. 

Virtually all classical novae (with the exception of the 10 known Galactic recurrent novae) exhibit ejection velocities in the range 300-3000 km/s (cf Warner (1995), Table5.2). Traveling at 3000 (300)  km/s, with no deceleration, the AT Cnc ejecta could have reached their present size in no less (no more) than 63 (630) yrs. 

The ejecta of novae suffer significant deceleration on a timescale of 50 - 100 years, as suggested 
by \citet{oor46} and directly measured by \citet{due87}. The observed expansion velocity dropped 
to half its initial value in 65, 58, 117 and 67 years, respectively, for the four classical novae 
V603 Aql, GK Per, V476 Cyg and DQ Her \citep{due87}. The observed peak ejection velocities of these four novae were 1700,
1200, 725 and 325 km/s, respectively. The remarkably small range in deceleration half-lifetimes t for the
four novae noted above are in excellent accord with the Oort snow-plough model's predictions. They demonstrate that AT Cnc's ejecta 
must have undergone at least one deceleration half-lifetime since its eruption. Thus the time since AT Cnc's last nova eruption 
is almost certainly double the lower limit noted above (i.e. 126 years) or greater. We cannot more strongly constrain the upper limit other than to say it is of order 1,000 years.

As a "sanity check" we note that GK Per (which erupted 111 years ago) is at the same distance as AT Cnc, and that it displays a shell almost half the size \citep{szd12} of the AT Cnc ejecta. GK Per will achieve AT Cnc's angular size in about 2 centuries, supporting our simple estimates of an age of a few centuries for AT Cnc.
 
The expansion rate of AT Cnc's ejecta should be directly measurable within a decade, providing a better-determined lower limit to its age. Measuring the deceleration will take longer, but its determination will enable the best possible estimate of the time elapsed 
since AT Cnc last erupted as a nova. This will, in turn, provide us with the first quantitatively measured estimate of the time 
required for an old nova to revert to its subsequent dwarf nova phase.
 
\section{Implications of the Existence of the AT Cnc Shell}

Our detection of a second nova shell surrounding a Z Cam-type dwarf nova further supports the claim that these stars are intimately connected with nova eruptions. There are 44 known and suspected Z Cam-type stars, as of April 2012; we have surveyed only half of them. An up-to-date list is maintained by Ringwald at https://sites.google.com/site/thezcamlist/the-list. If Z Cam were the only dwarf nova with a detected nova shell then one could have argued that we fortuitously captured a very rare and transient event. The existence of the AT Cnc old nova ejecta argues against that interpretation. We will soon announce a third Z Cam star surrounded by ejecta, further supporting our claim that old novae and Z Cam stars are intimately connected: strong evidence supporting hibernation.   

\section{Summary and Conclusions}

We report the optical narrowband detection of fragmented rings, 3 arcmin in diameter, surrounding the dwarf nova AT Cancri. The shell geometry is suggestive of bipolar, conical ejection seen nearly pole-on. A spectrum of the brightest part of the AT Cnc ejecta is similar to that of the ejecta of the classical nova GK Per, and of Z Cam, dominated by [NII], [OII] and [OIII] emission. The ejecta must be shock ionized. Galex FUV imagery reveals a similar-sized, FUV-emitting shell. We determine that AT Cnc is about 460 pc from Earth, with a system luminosity at maximum brightness that is $\sim  L_\sun$. The 1.5 arcmin radius of the AT Cnc ejecta corresponds to 0.2 pc at that distance, with a maximum shell mass of  $\sim  5 \times 10^{-5} M_\sun$, in excellent agreement with theoretical predictions of nova ejecta masses.

\clearpage
 
\begin{figure}
\figurenum{1}
\epsscale{1.0}
\plotone{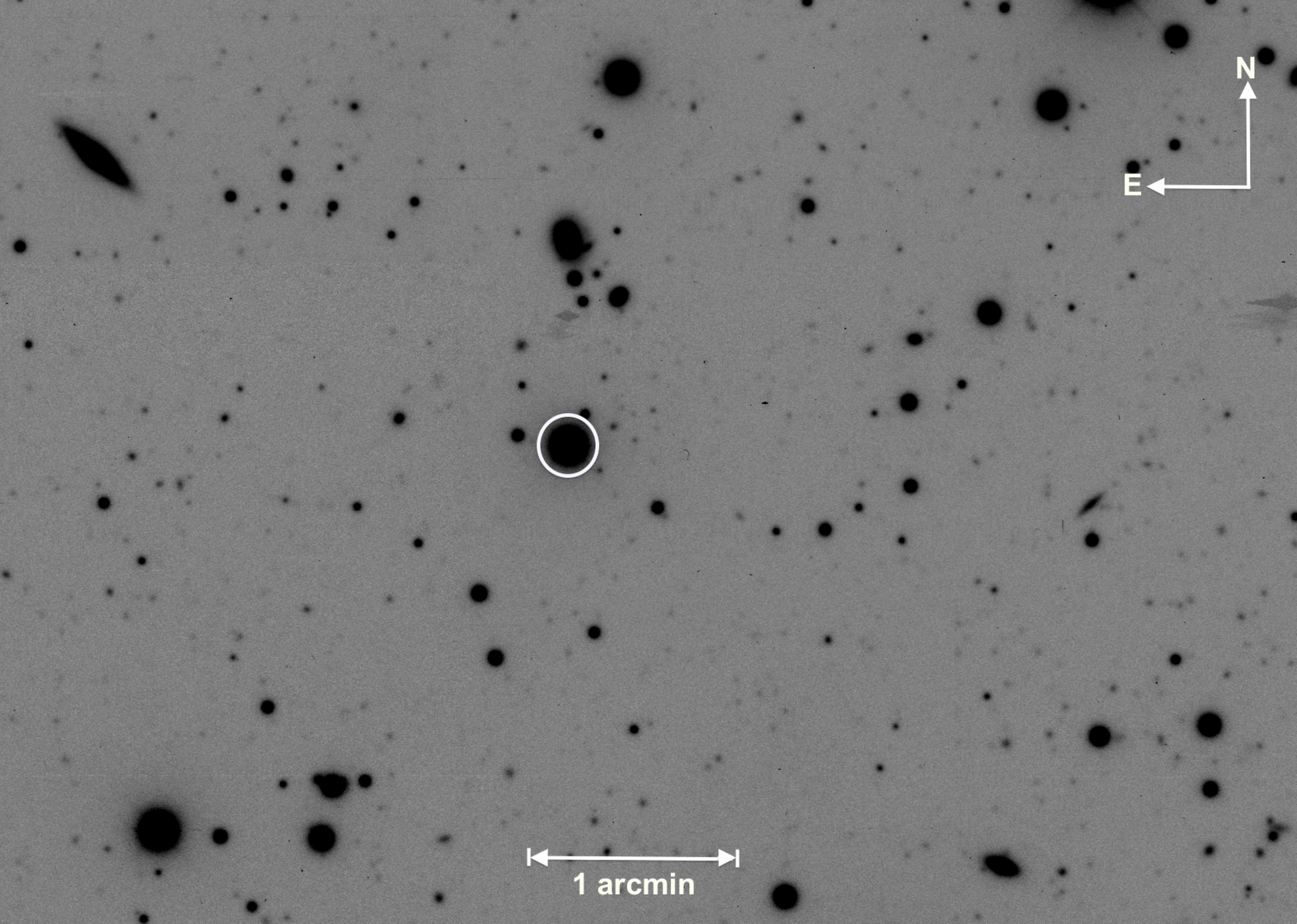}
\caption{A deep R band image of the field surrounding the Z Cam-type dwarf nova AT Cnc. AT Cnc is circled.}
\end{figure}

\clearpage
\begin{figure}
\figurenum{2}
\epsscale{1.0}
\plotone{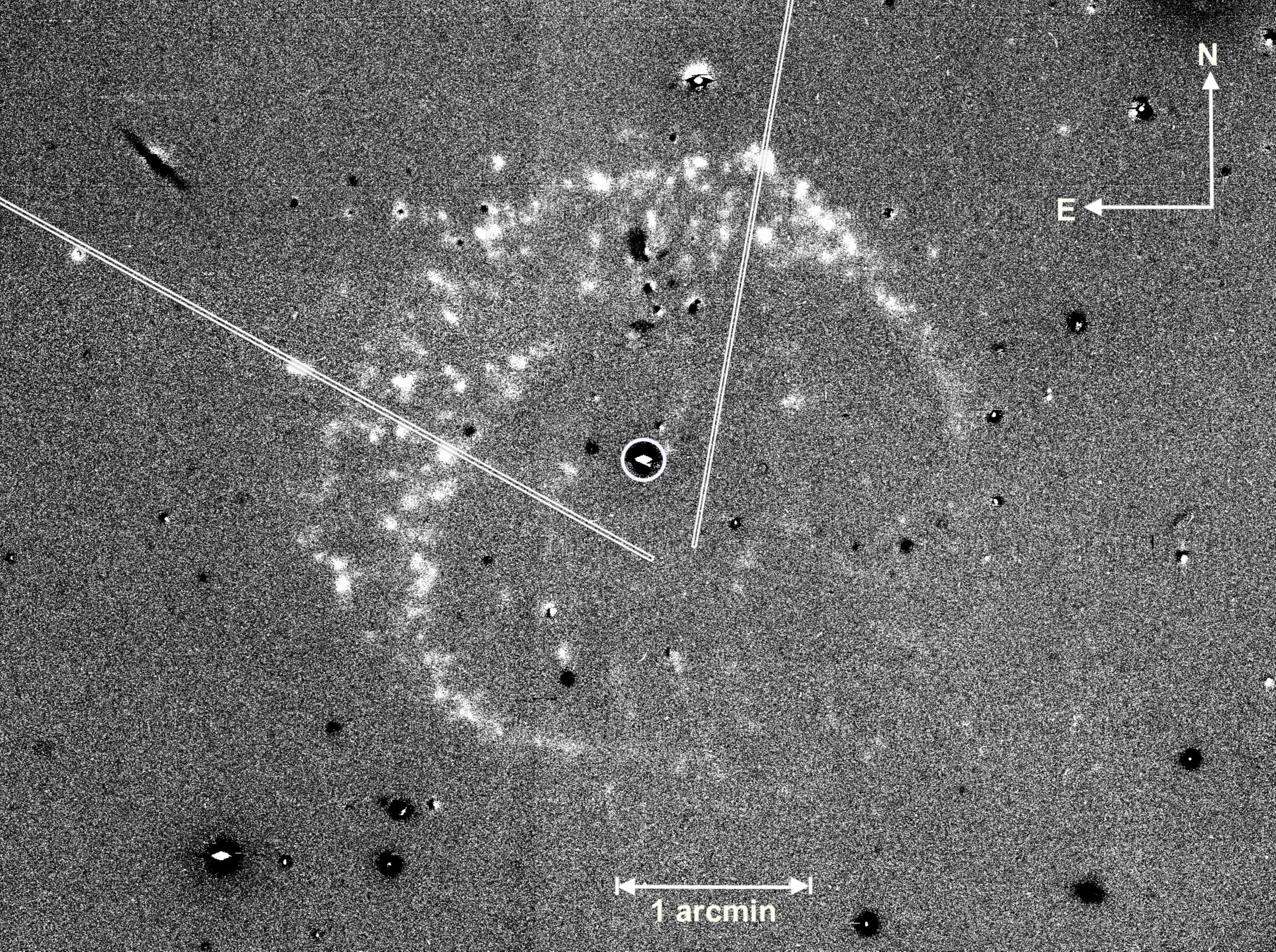}
\caption{A net H$\alpha$ +[NII] image of AT Cnc. AT Cnc is circled. The positions of two slits used to obtain spectra of the AT Cnc ejecta
are superposed on the image.}
\end{figure}

\clearpage

\clearpage
\begin{figure}
\figurenum{3}
\epsscale{1.0}
\plotone{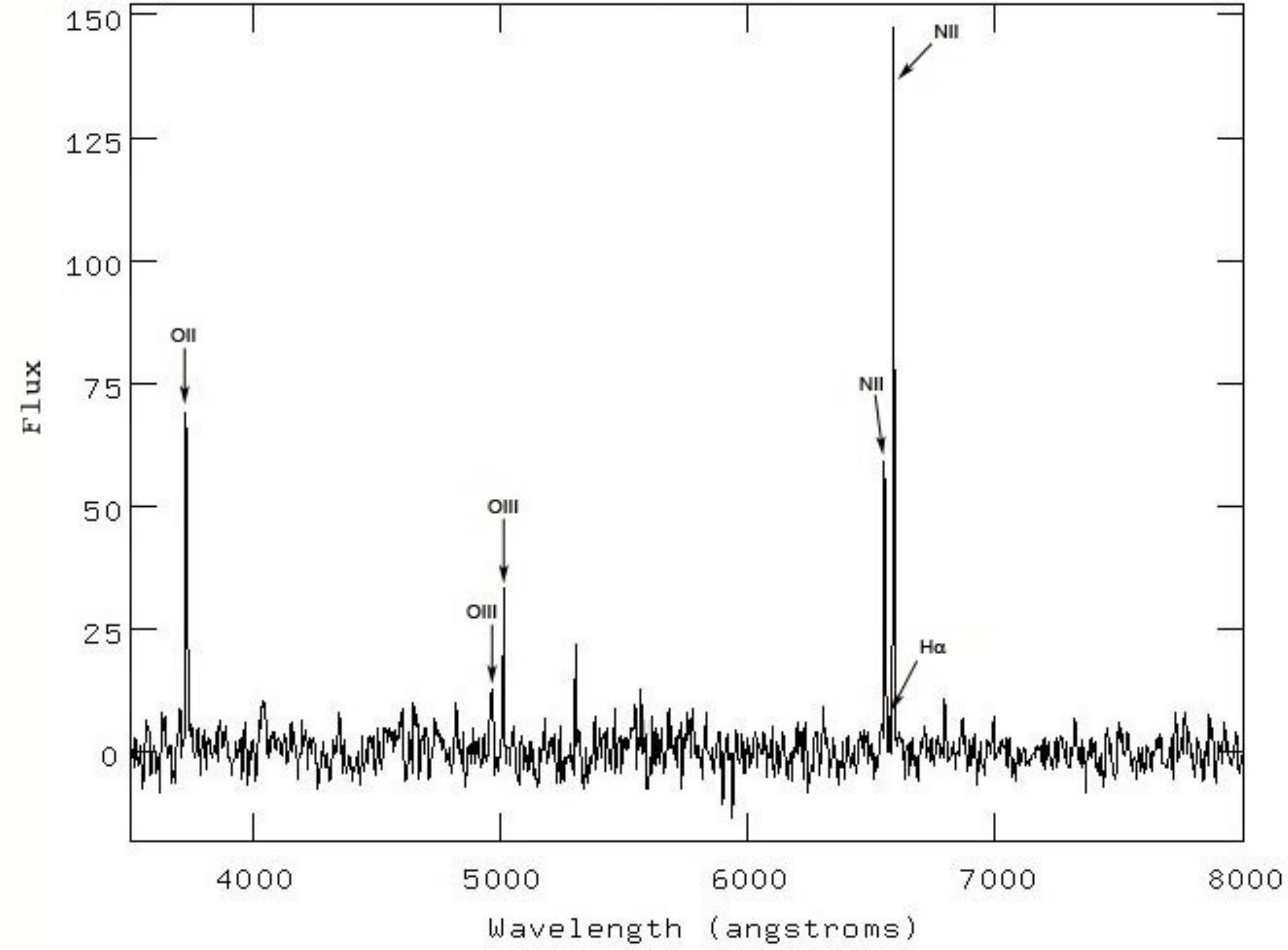}
\caption{The optical spectrum of the brightest knot in the ejecta of AT Cnc.}
\end{figure}

\clearpage

\clearpage
\begin{figure}
\figurenum{4}
\epsscale{1.0}
\plotone{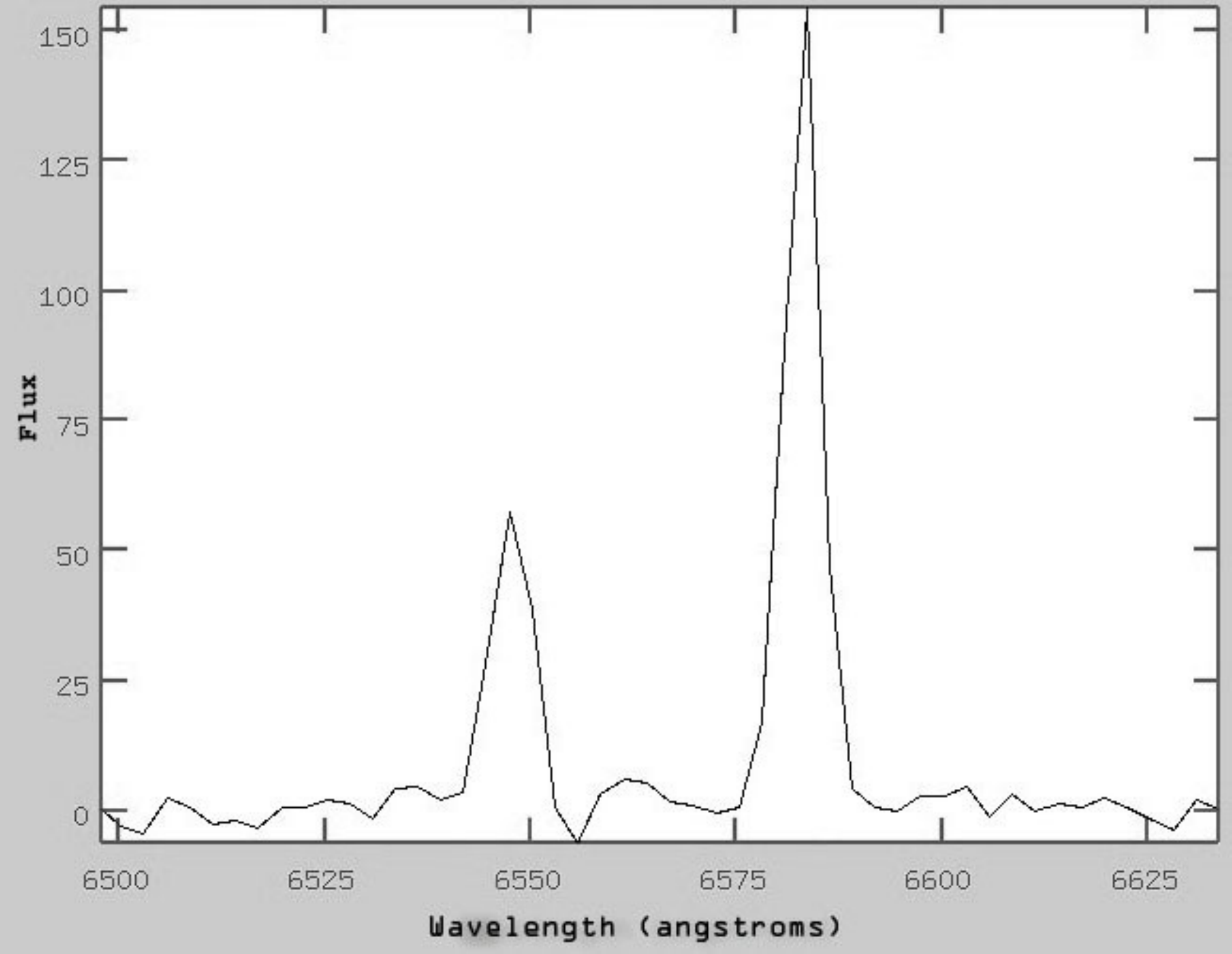}
\caption{A closeup of the optical spectrum of the AT Cnc ejecta, centered on the [NII] lines.}
\end{figure}

\clearpage
\begin{figure}
\figurenum{5}
\epsscale{1.0}
\plotone{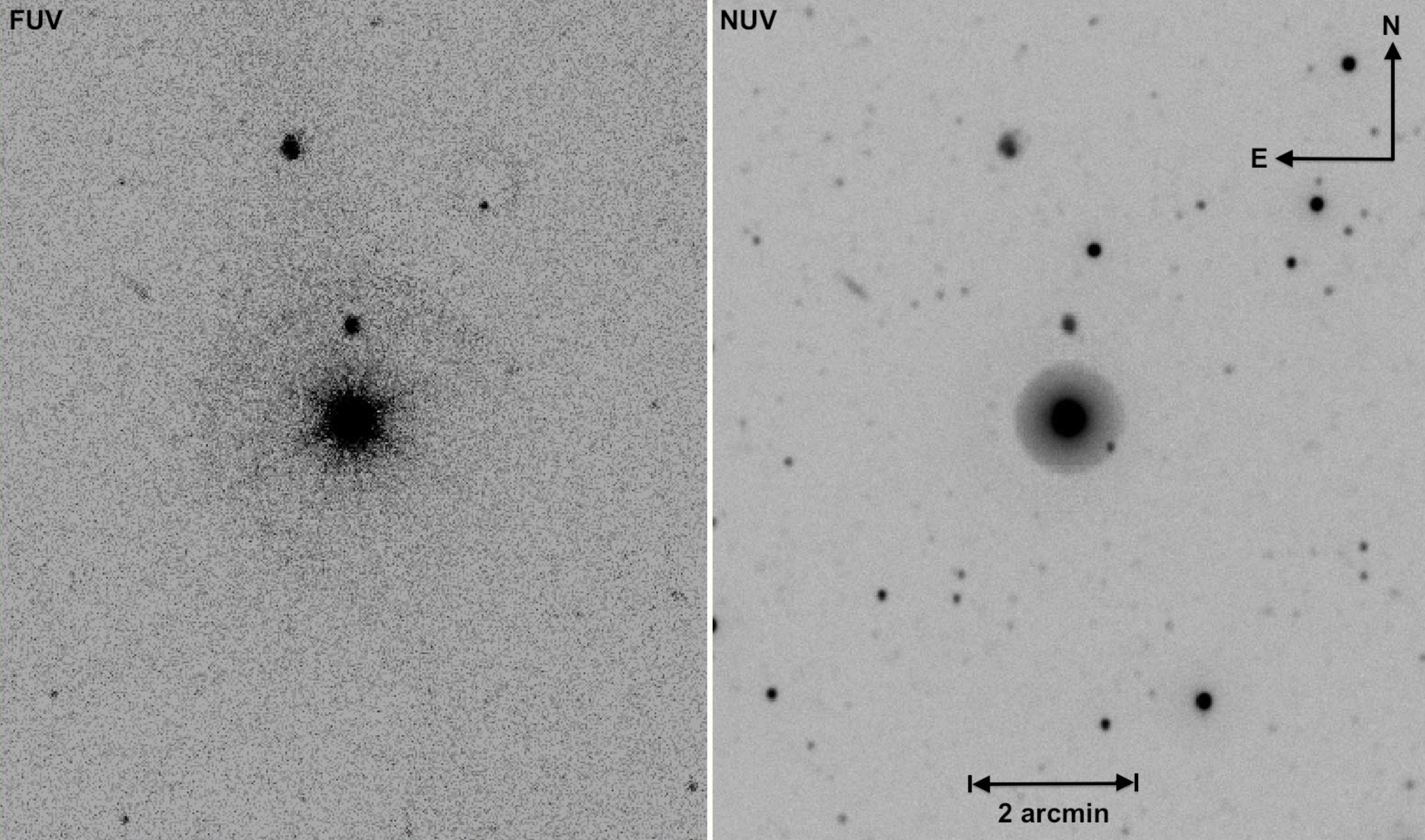}
\caption{The area around AT Cnc in the GALEX FUV and NUV images. The "halo" surrounding AT Cnc in the NUV image is instrumental, but the faint, extended FUV emission corresponds to the ejecta seen in Figure 2.}
\end{figure}

\clearpage

\clearpage
\begin{figure}
\figurenum{6}
\epsscale{1.0}
\plotone{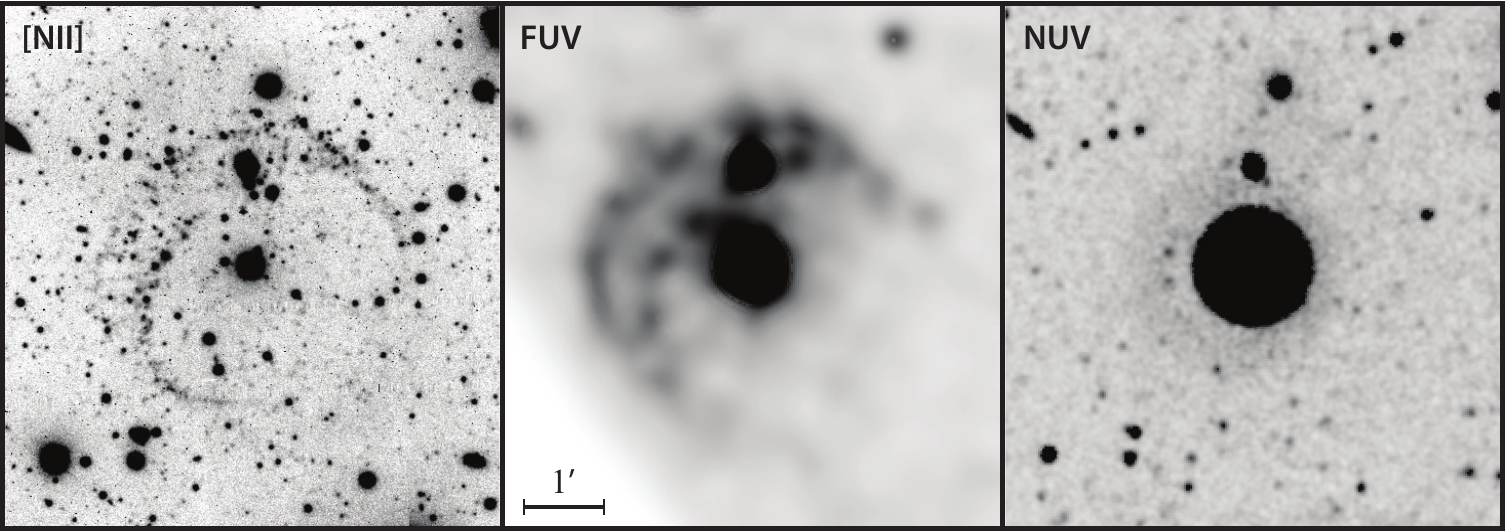}
\caption{Comparison of the optical narrowband [NII] image of AT Cnc with the GALEX FUV and NUV images to demonstrate the correspondence between optical narrowband [NII] and FUV emission.}
\end{figure}

\clearpage

\acknowledgments

GALEX (Galaxy Evolution Explorer) is a NASA Small Explorer, launched in April 2003. We gratefully acknowledge NASA's support for construction, operation, and science analysis for the GALEX mission. 

MMS gratefully acknowledges helpful conversations about AT Cnc and dwarf nova shells with Howard Bond and Christian Knigge.
  
\end{document}